\documentstyle[manuscript,aps,prb]{revtex}
\def\be{\begin{equation}}
\def\ee{\end{equation}}
\def\bea{\begin{eqnarray}}
\def\eea{\end{eqnarray}}
\begin{document}
\title{Linear Differential Equations and Orthogonal Polynomials: \\
A Novel Approach}
\author{N. Gurappa$^1$ \thanks{gurappa@ipno.in2p3.fr}, Prasanta K.
Panigrahi$^2$
\thanks{panisp@uohyd.ernet.in}
and T. Shreecharan$^2$ \thanks{panisprs@uohyd.ernet.in} .}
\address{ $^1$Laboratoire de Physique Theorique et Modeles Statistiques,
bat. 100,\\
Universit${\acute e}$ Paris-Sud-91405, Orsay, FRANCE;\\
$^2$School of Physics, University of Hyderabad, Hyderabad,\\
Andhra Pradesh, 500 046 INDIA.}
\maketitle
\begin{abstract}
A novel method, connecting the space of solutions of a linear
differential equation, of arbitrary order, to the space of
monomials, is used for exploring the algebraic structure of the
solution space. Apart from yielding new expressions for the
solutions of the known differential equations, the procedure
enables one to derive various properties of the orthogonal
polynomials and functions, in a unified manner. The method of
generalization of the present approach to the multi-variate case
is pointed out and also its connection with the well-known
factorization technique. It is shown that, the generating
functions and Rodriguez formulae emerge naturally in this method.
\end{abstract}
\draft

\newpage

{\section {Introduction}}

It has been recently shown, by two of the present authors
\cite{pani1}, that the solution of a linear differential equation
(DE), of an arbitrary order, can be mapped to the space of
monomials, if the operators, relevant to a given DE, can be
separated into a part containing the Euler operator ($D \equiv x
d/dx$) and the constants and another one retaining the other
operators. Separating a given DE into two different parts and
generating a series solution, from the solution of the simpler
one, by integration has been tried earlier in the literature
\cite{adomian}. The advantage of the present method derives from
the fact that, the Euler operator is diagonal in the space of
monomials and any other differential operator or monomial $(O ^d)$
is characterized by a definite degree $d$, with respect to the
Euler operator i.e., $[D, O^d] = dO^d$. By a judicious use of
these two results, the space of solutions of the above DE of
arbitrary order, is directly connected to the space of monomials,
avoiding any explicit integration, unlike the previous approaches.
This technique, not only yields novel expressions for both
polynomial and functional solutions of the known DEs, but can also
be straightforwardly extended to a wide class of multi-variate
DEs. It can be used to find solutions of equations \cite{prb1},
involving the so called Dunkl derivatives \cite{dunkl}; these type
of equations are being extensively studied in the current physics
and mathematics literature \cite{tom}. Furthermore, the approach
leads to the diagonalization of various many-body interacting
systems \cite{pani1,prb1,rapid} of the Calogero-Sutherland type
\cite{calo}.

In the present paper, we confine ourselves, to the single variable
case and explore the origin of various algebraic structures in the
space of solutions of the DEs and only briefly outline the
procedure to generalize this approach to the many-variable cases.
These algebraic structures manifest transparently here, since they
appear naturally in the space of monomials and our approach
connects these monomials to the solution space. It is worth
mentioning that these algebras are responsible for symmetries and
various degeneracies of physical problems \cite{gursey},
associated with the DE under consideration.

The paper is organized as follows. In the subsequent section, we
briefly outline the essential steps of our method and use the same
to obtain novel expressions for the solutions of the well-known
hypergeometric and confluent hypergeometric equations and also
show its applicability to generalized hypergeometric equations.
The ladder operators and the underlying algebras are then obtained
in section III. We then establish the connection of this approach
with the factorization technique and Supersymmetric quantum
mechanics (SUSY-QM). The utility of the method, in finding various
properties of the orthogonal polynomials, is then demonstrated in
section IV, by finding the generating functions and the Rodriguez
formulae, for a number of cases. We conclude in section V, after
pointing out a number of problems, where the present method can be
profitably employed.

{\section {Mapping between the solution space of differential
equations and monomials}}

In this section, we reproduce, for the sake of completeness, a
recently proposed method of solving linear differential equations
of arbitrary order \cite{pani1} and explicate the same, with a
number of examples.

After suitable manipulations, if a DE can be cast in the form \be
\label{ie} \left[F(D) + P(x,d/dx)\right] y(x) = 0 \quad, \ee
where, $D \equiv x \frac{d}{dx}$, $F(D) = \sum_{n = - \infty}^{n =
\infty} a_n D^n $, $a_n$'s are some parameters and $P(x,d/dx)$ is
a function of $x$, $d/dx$ and other operators, then the solution
to the DE can be written as, \be \label{an} y(x) = C_\lambda \left
\{\sum_{m = 0}^{\infty} (-1)^m
\left[\frac{1}{F(D)}P(x,d/dx)\right]^m \right \} x^\lambda \ee
provided, $F(D)  x^\lambda = 0$. Here $C_\lambda$ is a constant.
The proof is straightforward and follows by direct substitution:
\bea &&\left[F(D) + P(x,d/dx)\right] \left\{\sum_{m = 0}^{\infty}
(-1)^m
\left[\frac{1}{F(D)}P(x,d/dx)\right]^m \right \} x^\lambda \nonumber\\
\nonumber\\
= F(D)
&& \left[1 + \frac{1}{F(D)}P(x,d/dx)\right]
\left \{\sum_{m = 0}^{\infty} (-1)^m
\left[\frac{1}{F(D)}P(x,d/dx)\right]^m \right \}
x^\lambda \nonumber  \\ \nonumber \\
= F(D)
&& \sum_{m = 0}^{\infty} (-1)^m \left[\frac{1}{F(D)}P(x,d/dx)\right]^m
x^\lambda
\nonumber\\
&& + F(D) \sum_{m = 0}^{\infty}(-1)^m \left[\frac{1}{F(D)}P(x,d/dx) \right
]^{m + 1}
x^\lambda \nonumber \\ \nonumber \\ = F(D)
&& x^\lambda - F(D) \sum_{m = 0}^{\infty} (-1)^m
\left[\frac{1}{F(D)}P(x,d/dx)\right]^{m + 1}
x^\lambda \nonumber\\
&& + F(D) \sum_{m = 0}^{\infty}(-1)^m
\left[\frac{1}{F(D)}P(x,d/dx) \right ]^{m + 1} x^\lambda \nonumber
\\= 0 \quad. \eea Eq. (\ref{an}) connects the solution of a given
DE to the space of monomials. It needs to be emphasized that, the
inverse of $F(D)$ is well defined in the above expression, since
it is diagonal in the space of monomials.

We illustrate the working of this method in the context of the
well-known hypergeometric differential equation (HGDE)
\cite{grad}, given by, \be \label{hyper} \left[x^2
\frac{d^2}{dx^2} + {\left(\alpha + \beta + 1 \right)x
\frac{d}{dx}} + \alpha\beta - x \frac{d^2}{dx^2} - \gamma
\frac{d}{dx} \right] F{(\alpha,\beta;\gamma;x)}= 0 \quad. \ee
Collecting the part containing the powers of the Euler operator
and the constants, one gets, $F(D) \, = \, (D+ \alpha)(D +
\beta)$, and the condition $F(D) x^\lambda \, = \,0$, gives,
$\lambda \, = \, -\alpha, -\beta$. The series solution is then \be
\label{shg} F{(\alpha,\beta;\gamma;x)} = C_{(\alpha,\beta)}
\left\{\sum_{m=0}^{\infty} (-1)^m \left[\frac{-
1}{(D+\alpha)(D+\beta)} \left(x \frac{d^2}{dx^2} + \gamma
\frac{d}{dx}\right)\right]^m \right\} x^{- (\alpha, \beta)} \quad.
\ee It is easily seen that, modulo an overall normalization
factor, the above solution, is a rearranged form of the well-known
hypergeometric series. Making use of the following representation
of $1/(D+\beta)$, \be \frac{1}{(D+\beta)} = {\int_0}^\infty ds \,
e^{-s(D+\beta)} \ee and \be [(D+\beta),(x \frac{d^2}{dx^2}+ \gamma
\frac{d}{dx})]= - \left(x \frac{d^2}{dx^2}+ \gamma \frac{d}{dx}
\right)\quad, \ee one can easily show that \be
\left[\frac{-1}{(D+\alpha)(D+\beta)} (x \frac{d^2}{dx^2}+ \gamma
\frac{d}{dx})\right]^m x^{- \beta} \, = \,
\left[\frac{1}{(D+\alpha)} ( x \frac{d^2}{dx^2}+ \gamma
\frac{d}{dx}) \right]^m\frac{ x^{- \beta}}{m!} \quad. \ee Choosing
the normalization constant to match the conventional definition,
we can write the normalized solution for the HGDE as, \be
F(\alpha,\beta;\gamma;x)= (-1)^{-\beta} { \frac{\Gamma ( \alpha -
\beta)\Gamma(\gamma)} { \Gamma (\gamma - \beta) \Gamma(\alpha)}
\exp{\left[\frac{-1}{\left(D+\alpha \right)} \left(x
\frac{d^2}{dx^2}+ \gamma \frac{d}{dx}\right) \right]}} \, . \,
x^{-\beta} \quad. \ee The exponential form is a novel expression
for $F(\alpha,\beta;\gamma;x)$, unknown in the literature, to the
best of the authors' knowledge. The choice $ \lambda = - \alpha$
will lead again to the hypergeometric series, as the series is
symmetric under the exchange of $\alpha$ and $\beta$. Since,
$\frac{1}{(D+\alpha)}(x d^2/dx^2 + \gamma d/dx)$ lowers the degree
of a given monomial by one, it can be seen that, when $-\beta$ is
an integer, the above is a polynomial solution of the HGDE.

Similarly, the normalized polynomial solution for the confluent
hypergeometric (CH) differential equation, \be \left[x
\frac{d^2}{dx^2}+ (\gamma - x) \frac{d}{dx} - \alpha
\right]\Phi(\alpha;\gamma;x) = 0 \quad, \ee is \be \label{chsol}
\Phi(\alpha;\gamma;x) = (-1)^{ -\alpha}
\frac{\Gamma(\gamma)}{\Gamma(\gamma-\alpha)} \exp{\left[- x
\frac{d^2}{dx^2}- \gamma \frac{d}{dx}\right]}\,.\, x^{- \alpha}
\quad. \ee

It is worth mentioning that, one can also modify the HGDE, by
multiplying it with $x$, which yields, $F(D) \equiv D(D+\gamma
-1)$ and $P(x,d/dx) \equiv -x(D+\alpha)(D+\beta)$. The
corresponding solution is \be F{(\alpha,\beta;\gamma;x)} =
C_{(0,1-\gamma)} \left\{\sum_{m=0}^{\infty} (-1)^m \left[\frac{
-1}{D(D+\gamma -1)}x(D+\alpha)(D+\beta)\right]^m \right\} x^{
(0,1-\gamma)} \quad, \ee since $F(D)x^\lambda = 0$ gives $\lambda
= 0,1-\gamma$. For $\lambda=0$, the series can be written in the
well-known form \be F{(\alpha,\beta;\gamma;x)} = \sum_{n=0}^\infty
\frac{(\alpha)_n (\beta)_n}{(\gamma)_n} \frac{x^n}{n!} \quad, \ee
where, \be a_n =  a(a + 1)(a + 2) \cdots (a + n - 1) \quad, \ee is
the Pochammer symbol. The corresponding exponential form is \be
F{(\alpha,\beta;\gamma;x)} = C_0 \exp
\left[x\frac{(D+\alpha)(D+\beta)}{(D+\gamma)}\right]\,.\,1
\quad.\ee The other choice, namely $\lambda = 1-\gamma$, yields
the other linearly independent solution. Analogous results follow
for CHDE. Hence, both the polynomial and the series solutions are
obtained by the present method.

The procedure for finding the solutions for CHDE and HGDE easily
extends to the generalized hypergeometric cases \cite{slater},
given by the equations of the type, \be [ \Theta(\Theta + b_1 - 1)
\cdots (\Theta + b_p - 1) - z(\Theta + a_1) \cdots (\Theta + a_p
+1)] y = 0 \quad, \ee where $\Theta \equiv z d/dz$, with $z$
being, in general, complex. The generalized hypergeometric series,
is denoted by $_{p+1} F_p$, where $p+1$ and $p$ are the number of
parameters appearing in the numerator and the denominator of the
generalized HG series, respectively. Since $\Theta$ is the Euler
operator and the DE is already separated into $F(\Theta)$ and
$P(z, d/dz)$, the solutions easily follow. We consider the case of
the $_3F_2$ series below, because of its importance in the quantum
theory of angular momentum \cite{srinivas}.

The DE for the $_3F_2$ series \be [ \Theta (\Theta + b_1 -
1)(\Theta + b_2 - 1) - z(\Theta + a_1)(\Theta + a_2)(\Theta +
a_3)] \, _3F_2 = 0 \quad,\ee yields $F(\Theta) \equiv
\Theta(\Theta + b_1 - 1)(\Theta + b_2 - 1)$ and $P(z, d/dz) \equiv
-z(\Theta + a_1) (\Theta + a_2)(\Theta + a_3)$. The solution is
then \be _3F_2 = C_{(0,1-b_1,1-b_2)} \left\{\sum_{m=0}^{\infty}
(-1)^m \left[\frac{1}{F(\Theta)} P(z, d/dz)\right]^m \right\}
z^{(0,1-b_1,1-b_2)} \quad, \ee where $ F(\Theta) z^\lambda = 0$,
has yielded three solutions, $\lambda = 0,1-b_1,1-b_2$. For
$\lambda = 0$, the above series can be expanded and rearranged in
the conventional form \be _3F_2= \sum_{n=0}^\infty \frac{(a_1)_n
(a_2)_n(a_3)_n}{(b_1)_n(b_2)_n} \frac{z^n}{n!} \quad. \ee Like the
previous examples, an exponential form for the $_3F_2$ can also be
written down easily.

The other roots, $1-b_1$ and $1-b_2$, give the other two linearly
independent solutions. It is clear from the expression for the
$_3F_2$ series, that it terminates, when either $a_1, a_2$ or
$a_3$, is a negative integer or zero.

Multiplying the above DE with $-1/z$, one gets, \be [(\Theta +
a_1)(\Theta + a_2)(\Theta + a_3)- \frac{d}{dz}(\Theta + b_1 -
1)(\Theta + b_2 - 1)] \, _3F_2 = 0 \quad, \ee where $F(\Theta)$
and $P(z, d/dz)$ are now given by $F(\Theta)=(\Theta + a_1)(\Theta
+ a_2)(\Theta + a_3)$ and $P(z, d/dz) = -(d/dz)(\Theta + b_1 -
1)(\Theta + b_2 - 1)$. Therefore, the solution, in this case,
turns out to be \be _3F_2 = C_{(a_1,a_2,a_3)}
\left\{\sum_{m=0}^{\infty} (-1)^m \left[\frac{1}{F(D)} P(z,
d/dz)\right]^m \right\} z^{-(a_3,a_2,a_1)} \quad, \ee where
$a_1,a_2,a_3$ have to be negative integers to yield a polynomial
solution.

By now, it is clear that for a number of DEs, $F(D) x^\lambda =
0$, leads to the linearly independent solutions, when the
solutions are nondegenerate. In cases, where this is not possible,
appropriate modification of the DE, has straightforwardly yielded
the linearly independent solutions. Below, we give two more
examples to illustrate these points. In particular, the second
example deals with the case, where $F(D) x^\lambda = 0$, leads to
degenerate solutions. These examples will also point out the
connection of the indicial equation, in the conventional series
solution method, to the present one. In the standard series
solution method \cite{morse}, for the DE \be 4x\frac{d^2y}{dx^2} +
2\frac{dy}{dx} + y = 0 \quad, \ee  the two distinct roots of the
{\em indicial equation} $c = 0,1/2$, lead to two linearly
independent solutions. Multiplication by $x$ yields \be
4x^2\frac{d^2y}{dx^2} + 2x\frac{dy}{dx} + xy = 0 \quad, \ee which
implies, $F(D) = 4x^2\frac{d^2}{dx^2} + 2x\frac{d}{dx} = 2D(2D-1)$
and $P(x,d/dx)=x$. Hence, $F(D)x^\lambda=0$ obtains $\lambda = 0,
1/2$, which provide the two linearly independent solutions, as is
obtained in the standard approach.

In our second example,\be x \frac{d^2}{dx^2} + \frac{dy}{dx} + y =
0 \quad,\ee the roots of the {\em indicial equation}, $c^2 = 0$
are degenerate. Multiplying the above DE with $x$, to bring it to
the form given by Eq. (\ref{ie}), one gets $(D^2 + x)y = 0$; hence
$F(D) = D^2$ and $P(x) = x$. It is clear that, $F(D)x^\lambda =
0$, also leads to the same degenerate case as obtained by the
method of series solution. In this scenario, one has to employ the
established methods for finding out the other linearly independent
solution \cite{morse}.

The procedure developed here is applicable to a wide range of
functions and polynomials. Some of the well-known ones, explicitly
checked by the authors are, Meijer's G-function, Struve, Lomel,
Anger, Weber, Bessel functions, Gegenbauer, Neumann's, Jacobi,
Schl\"afli, Whittaker and Chebyshev polynomials. It is also
applicable in the periodic cases. For example, solution of DE with
the following periodic potential
\begin{eqnarray} \label{pp} \frac{d^2 y}{d x^2} + a \cos(x) y = 0
\quad,
\end{eqnarray}
can be found, after multiplying Eq. (\ref{pp}) by $x^2$ and
rewriting $x^2 \frac{d^2}{d x^2}$ as $(D - 1) D$,
\begin{eqnarray}
y(x) = \sum_{m , \{n_i\} = 0}^{\infty} \frac{(- a)^m}{m!}
\left\{\prod_{i = 1}^m \frac{(-1)^{n_i}}{(2n_i)!} \right\}&&
\left\{\prod_{r = 1}^m \frac{(2 [m + \lambda/2 - r +
\sum_{i=1}^{m+1-r}n_i])!} {(2 [m + \lambda/2 + 1 - r +
\sum_{i=1}^{m+1-r}n_i])!}
\right\} \nonumber\\
&& \times \,\,x^{2(m + \sum_{i=1}^m n_i + \lambda/2)} \quad.
\end{eqnarray}
Here, $\lambda = 0$ or $1$. In the same manner, one can write down
the solutions for the Mathieu's equation as well \cite{grad}.

Although this paper is devoted to the single variable case, it is
worth pointing out that the generalization of this method to the
many-variable case is immediate \cite{pani1,prb1}. This can be
accomplished by denoting $\bar D = \sum_i D_i \equiv \sum _i x_i
\frac{d}{dx_i}$, where $i = 1, 2, \cdots N$. Using the fact,
$F(\bar D) X^\lambda = 0$ has solutions, in the space of monomial
symmetric functions \cite{sf}, a number of many-body equations can
be solved, in a manner analogous to the single variable case.

{\section {Algebraic structure of the solution space}}

We now proceed to study the algebraic properties of the space of
solutions. The advantages of the present approach, as compared to
the previous ones \cite{gursey}, lie in the following two facts.
First of all, {\em a priori}, no symmetry of the DE is assumed.
Secondly, the ladder operators are straightforward to construct in
the space of the monomials, which can be brought to the space of
solutions, via a similarity transformation, with the aid of the
exponential form of the solutions. The only criterion in the
choice of the ladder operators, in the space of monomials, is
that, after the similarity transformation, the resulting
operators, are well-defined; these operators then yield the
symmetry algebra.

Keeping in mind, the appearance of CHDE and HGDE in diverse
physical systems, we explicitly work out the generators of the
symmetry algebras, for these two cases, for the corresponding
polynomial solutions. Application to other DEs can be carried out
in a similar manner.

The solution of the CHDE,\be \label{chsol} \Phi(\alpha;\gamma;x) =
(-1)^{ -\alpha} \frac{\Gamma(\gamma)}{\Gamma(\gamma-\alpha)}
\exp{\left[- x \frac{d^2}{dx^2} - \gamma \frac{d}{dx}\right]}\,.\,
x^{- \alpha} \quad, \ee leads to a polynomial solution, only if,
$\alpha$ is a negative integer $(-n)$ or zero. The above form of
the solution, immediately suggests the lowering operator to be,
\be J_{-} \equiv \left(x \frac{d^2}{dx^2}+ \gamma
\frac{d}{dx}\right) \quad, \ee since $J_-$ commutes with the
exponential and hence can lead to a lowering operator at the level
of the polynomial. $J_-$ reduces the degree of the polynomial by
one, \be J_- e^{-J_-}x^n = n(\gamma+n-1)e^{-J_-}x^{n-1} \quad. \ee
After taking into account the normalization factors, we get
\be\label{clo} \left( x \frac{d^2}{dx^2}+\gamma
\frac{d}{dx}\right) \Phi(-n;\gamma;x) = -n \Phi(-n+1;\gamma;x)
\quad. \ee Choosing $J_+ \equiv x$, as the raising operator at the
level of the monomials and introducing an identity operator, in
the following manner, \be
e^{-J_-}xx^n=e^{-J_-}xe^{+J_-}e^{-J_-}x^n \quad, \ee one gets, \be
\label{cra} \left[x - 2x\frac{d}{dx} - \gamma + x \frac{d^2}{dx^2}
+ \gamma \frac{d}{dx}\right] \Phi(-n;\gamma;x) =  -(n+\gamma)
\Phi(-n-1;\gamma;x)  \quad. \ee The similarity transformation of
$x$, the raising operator, at the level of the monomials, led to
the raising operator at the level of the polynomials. It can be
easily seen that, $[J_+, J_-] = -2J_0$, where $J_0 = D + \gamma/2$
and \bea \left[{J_0},{J_\pm }\right] = \pm J_\pm \quad. \eea This
is the well-known $SU(1,1)$ dynamical algebra, in the solution
space of the CHDE. With an appropriate choice of the measure, the
operators $J_+$ and $J_-$ can be made the formal adjoints of each
other. It is worth pointing out that, the simplest choice $d/dx$
as the lowering operator for the monomials, does not lead to a
convergent expression at the level of the polynomials. It should
be pointed out that other forms of ladder operators can also be
found.

A straightforward calculation, taking $x$ and $d/dx$ as the
raising and lowering operators at the level of the monomials,
leads to the Heisenberg algebra, $[a,a^\dagger]=1$, for the
Hermite polynomials\be H_n(x) =
C_n\exp(-\frac{1}{4}\frac{d^2}{dx^2})\,.\,x^n \quad.\ee The
exponential form of the solution of HGDE suggests the simplest
lowering operator to be \be {\tilde J}_- = \frac{1}{(D+\alpha)}
\left(x \frac {d^2}{dx^2}+ \gamma \frac{d}{dx}\right) \equiv
{\tilde T} J_- \quad,\ee which lowers the degree of the monomials
by one. Here ${\tilde T} = 1/(D+\alpha)$ and $ J_- = [x
(d^2/dx^2)+ \gamma (d/dx)]$. It is to be noted that $J_-$ can also
act as the lowering operator at the level of the monomials;
however, the latter choice will not lead to a convergent
expression in the solution space.

Operating ${\tilde J}_-$ on $x^n$, one gets, \be {\tilde J}_- x^n=
\frac{n(\gamma+n-1)}{(\alpha+n-1)}x^{n-1} \quad, \ee which
straightforwardly extends to, \be \frac{1}{(D+\alpha)}\left(x
\frac{d^2}{dx^2}+ \gamma \frac{d}{dx} \right)
F(\alpha,-n;\gamma;x) = -n F(\alpha;-n+1;\gamma;x) \quad,\ee at
the level of the polynomials. The raising operator, at the level
of the monomials, needs to be chosen carefully due to the presence
of ${\tilde J}_-$ in the exponential.

At this point, we note that, the existence of a canonical
conjugate operator ${\tilde J}_+$ for ${\tilde J}_-$, such that,
$[{\tilde J}_-,{\tilde J}_+] = 1$, akin to the Heisenberg algebra,
would imply \be e^{{\tilde J}_-} {\tilde J}_+ x^n = ({\tilde J}_+
- 1) e^{{\tilde J}_-} x^n \quad.\ee This suggests ${\tilde J}_+$
can be used as a raising operator in the space of monomials, which
can be written in a compact form in the space of solutions of the
HGDE.

The procedure for finding the canonical conjugate is
straightforward \cite{shanta}. We briefly outline the same below
and use it for finding ${\tilde J}_+$. Although we have used this
procedure for convenience, other forms of raising operators are
possible to construct. The method for obtaining them will be
illustrated below. Denoting $J_0 \equiv xd/dx + \gamma/2$, one
finds $[J_0,{\tilde J_-}]=-{\tilde J_-}$. Starting with a function
$T(J_0)$, whose required properties will become clear in the
subsequent steps, we define \bea \nonumber {\tilde J_+} &=& x
T(J_0) \\ &\equiv& J_+ T(J_0) \quad. \eea It is easy to check
that,\bea [J_0,J_+]= J_+ \quad{\mathrm and}\quad [J_+, J_-]=-2J_0
\quad. \eea Since $-2J_0 = g(J_0)-g(J_0-1)$, with $g(J_0)= -
J_0(J_0+1)$, the Casimir operator $C$ commuting with all the three
generators can be written as, $C=J_-J_+ + g(J_0) = J_+J_- + g(J_0
- 1)$ \cite{rocek}. Starting from $[{\tilde J_-},{\tilde J}_+
]=1$, one finds \bea \nonumber {\tilde T(J_0)} T(J_0){J_-}{J_+} -
T({J_0-1}){\tilde T}({J_0-1}){J_+}{J_-}=1 \quad,\eea which leads
to \bea\nonumber T(J_0)=[{\tilde T}(J_0)]^{-1} \frac{J_0 +
\delta}{[C-g(J_0)]} \quad. \eea Here $\delta$ is an arbitrary
constant. Simplification yields \bea \nonumber {\tilde J}_+ =
\frac{ \left(J_0 + \alpha-1 \right) \left(J_0 + \delta -1 \right)}
{\left(J_0 + \gamma-1 \right) {J_0}} x \quad. \eea Demanding that
$[{\tilde J}_-,{\tilde J}_+] = 1$, holds on the lowest monomial
i.e., $1$, one obtains $\delta = 1$. This leads to \be {\tilde
J}_+= \frac{ \left( {J_0}+\alpha-1 \right)}
{\left({J_0}+\gamma-1\right)}{J_+} \quad. \ee Now the raising
operator in the solution space can be obtained via a similarity
transformation: \be e^{-{\tilde J}_-}{\tilde J}_+ e^{{\tilde
J}_-}\,e^{-{\tilde J}_-}x^n= \frac{\left( \alpha + n \right)}
{\left(\gamma + n \right)} e^{-{\tilde J}_-}x^{n+1} \quad. \ee The
above expression after restoring the normalization factors
yields,\be \left[ 1 - \frac{(J_0 + \alpha-1)}{(J_0+\gamma-1)}x
\right] F(\alpha;-n;\gamma;x)=F(\alpha;-n-1;\gamma;x)\quad. \ee

\noindent As noted earlier, the raising operator obtained above is
not unique, one can construct other raising operators; some
examples are \be {\bar J_+} = (x+x^2 \frac{d}{dx}) \quad {\mathrm
and} \quad {\hat J}_+ = (x+x^2 \frac{d}{dx}) {\hat T(C,{\tilde
J_0})} \quad. \ee For the former case, explicit computation leads
to \be \left[1- \frac{(D + \alpha - 1)}{(D + \gamma - 1)D}(x + x^2
\frac{d}{dx}) \right] F(\alpha; -n; \gamma; x) = F(\alpha; -n-1;
\gamma; x) \quad, \ee It is interesting to note that, the algebra
satisfied by ${\bar J_+},J_-$ and $D$ is a quadratic algebra
\cite{skly,vsk}, since \be [J_+,J_-]=-2(\gamma + 1/2)D - 3 D^2 -
\gamma \quad, \ee and \bea \nonumber [D, J_{\pm}] = \pm
J_{\pm}.\eea The algebra satisfied by $J_+, J_-$ and $J_0$ is the
well known $SU(1,1)$ algebra. Here $J_+, J_-$ can be made formal
adjoints of each other with the appropriate choice of the measure.
The point to note is that, both the above mentioned $SU(1,1)$ and
the quadratically deformed algebra, have the solution space of
HGDE, as their irreducible representations.

The above ladder operators can be suitably rearranged to yield the
ladder operators obtained by the factorization method (FM)
\cite{sch,hull} and SUSY-QM \cite{k2}. These ladder operators can
be used for obtaining wavefunctions of a number of quantum
mechanical problems \cite{k2,k1,jaf}, after appropriate measures
are introduced. We take the well-known example of the Laguerre DE,
for the purpose of establishing the above mentioned connection:
\be \left[ x \frac{d^2}{dx^2}+ (\alpha + 1 -x )\frac{d}{dx}+ n
\right]L_n^\alpha =0 \quad. \ee The solution of the above DE is
the familiar Laguerre polynomial, \be L^{\alpha}_n =
\frac{(-1)^n}{n!} \exp \left[- x \frac{d^2}{dx^2} - (\alpha+1)
\frac{d}{dx} \right]\,.\,x^n\quad. \ee In order to find the
raising operator for $L_n^\alpha$, analogous to the ones obtained
by FM and SUSY-QM, we start with \bea\nonumber \left[x+x
\frac{d}{dx}-n \right]x^n = x^{n+1}\quad,\eea as the raising
operator at the level of the monomials. Following the method
employed for CHDE, we get \be \label{rec1} \left[x
\frac{d}{dx}+(\alpha+1)+ n -x \right] L^{\alpha}_n =
(n+1)L^{\alpha}_{n+1}\quad. \ee This can be cast in the form\be
\label{lag} L^{\alpha}_n = \prod_{k=1}^n \left[x \frac{d}{dx}+
\alpha+k - x \right]1\quad,\ee or \bea\nonumber L^{\alpha}_n =
A^\dagger(\alpha+1) A^\dagger(\alpha+2)\cdots A^\dagger(\alpha+n)1
\quad,\eea where $A^\dagger(\alpha+n) \equiv \left[x \frac{d}{dx}
+ \alpha + n - x \right]$. These shifted operators have found
application in the construction of coherent states
\cite{benedict}. This form of the polynomial, modulo
normalizations; matches with the ones obtained from FM and
SUSY-QM. Using the raising operator for CHDE and HGDE, one can
al;so obtain similar expressions for other polynomials.

It should be emphasized that FM and SUSY-QM, have taken recourse
to a special property of the equations under study called shape
invariance \cite{ged} for arriving at these results. Furthermore
it was necessary to introduce a host of intermediate systems. In
contrast the present technique does not presume any such special
properties of the DE.

Since we have already provided the raising operator, for the sake
of completeness, we derive the lowering operator for $L_n^\alpha$
in a convenient form. Starting from, \bea\nonumber \left[x
\frac{d}{dx}-n+x \frac{d^2}{dx^2}+(\alpha+1) \frac{d}{dx}\right]
x^n = n(n+\alpha)x^{n-1}\quad\eea a suitable similarity
transformation yields \be \label{rec2} \left[x \frac{d}{dx}-n
\right]L^{\alpha}_n = -(n+\alpha)L^{\alpha}_{n-1} \quad. \ee Eq.
(\ref{rec1}) and Eq. (\ref{rec2}) are the standard recurrence
relations \cite{grad} for the Laguerre polynomials. We can also
derive those operators which change the values of $\alpha$. The
following steps lead to, \bea \frac{d}{dx}L^{\alpha}_n(x) &=&
\frac{(-1)^n}{n!} \frac{d}{dx} \exp\left[-x \frac{d^2}{dx^2} -
(\alpha+1) \frac{d}{dx}\right]x^n
\\ \nonumber \\ \nonumber &=&\frac{(-1)^n}{n!} \exp\left[-
\frac{d}{dx}x\frac{d}{dx} -
(\alpha+1)\right]\frac{d}{dx}x^n \\
\nonumber \\  &=& - L^{\alpha+1}_{n-1}(x) \quad.\eea Similarly for
HG series we get, \bea\nonumber
\frac{d}{dx}F(\alpha,\beta;\gamma;x)&=&
 -\beta(-1)^{-\beta}
\frac{\Gamma ( \alpha - \beta)\Gamma(\gamma)} { \Gamma (\gamma -
\beta) \Gamma(\alpha)} \times
\\ &\times& \exp{\left[\frac{-1}{\left(D+\alpha + 1 \right)}
\left(x \frac{d^2}{dx^2}+ (\gamma+1) \frac{d}{dx}\right)
\right]}\, . \, x^{-\beta - 1} \\ \nonumber \\ &=& \frac{\alpha
\beta}{\gamma} F(\alpha+1,\beta+1;\gamma+1;x)\eea

\section{Appplication to Classical Orthogonal Polynomials}

{\subsection{Rodriguez Formula}}

In this section, we elaborate on the applicability of the approach
developed here for finding other properties of the solution space.
We start with the Rodriguez formula (RF) and show how these can be
obtained, with the examples of Laguerre and Hermite polynomials.

\noindent \underline{{\bf RF for Laguerre polynomials}}

For simplicity, we consider the expression of $L^0_n$, \be
L^{0}_n(x)= \frac{(-1)^n}{n!} \exp \left(- x \frac{d^2}{dx^2} -
\frac{d}{dx} \right)x^n \quad. \ee Defining $B \equiv xd^2/dx^2
+d/dx$ and introducing an identity operator, we get \bea \nonumber
L^{0}_n(x)= \frac{(-1)^n}{n!} e^{-B}x^ne^{B}e^{-B}\,1 \quad. \eea
The above expression simplifies to \be \label{rf1} L^{0}_n(x)=
\frac{(-1)^n}{n!} \left[x-2x
\frac{d}{dx}-1+x\frac{d^2}{dx^2}+\frac{d}{dx}\right]^n e^{-B}\,1
\quad,\ee which can be written in the form \bea\nonumber
L^{0}_n(x)= \frac{(-1)^n}{n!} e^{x}e^{-x}\left[x-2x
\frac{d}{dx}-1+x\frac{d^2}{dx^2}+\frac{d}{dx}\right]^n
e^{x}e^{-x}\,1 \quad. \eea Further simplification yields,\be
L^{0}_n(x)= \frac{(-1)^n}{n!} e^{x}\left[\frac{d}{dx}x\frac{d}{dx}
\right]^ne^{-x}\quad, \ee and hence \be L^{0}_n(x)=
\frac{1}{n!}e^x \frac{d^n}{dx^n}(e^{-x}x^n)\quad. \ee This is the
well-known Rodriguez formula for the Laguerre polynomials.

\noindent {\underline {\bf RF for the Hermite polynomials}}

The normalized solution of the Hermite DE is given by,
\be\label{h} H_n(x)= 2^n \exp(-\frac{1}{4} \frac{d^2}{dx^2})\,x^n
\quad. \ee Denoting $A \equiv \frac{1}{4} \frac{d^2}{dx^2}$ and
writing \be H_n(x)= 2^n e^{-A} x^n e^{A}e^{-A}\,1 \quad, \ee we
obtain \be \label{rf2} H_n(x)\,=\,\left[2x-\frac{d}{dx}\right]^n
e^{-A}\,1 \quad. \ee Introducing an identity operator in the form,
\be H_n(x)\,=\,e^{x^2}e^{-x^2} \left[2x-\frac{d}{dx}\right]^n
e^{x^2}e^{-x^2} e^{-A}\,1 \ee one obtains the RF as \be H_n (x)=
(-1)^n e^{x^2} \frac{d^2}{dx^2} e^{-x^2} \quad. \ee

Eq. (\ref{rf1}) and Eq. (\ref{rf2}) reveal that simplification of
the raising operators by introducing appropriate identity
operators led to the RF. Using the raising operators obtained in
the previous section, one can easily extend these results to other
polynomials.

One can also start from the RF and obtain Eq.(\ref{h}).
Explicitly,\bea \nonumber
H_n (x)&=&(-1)^ne^{x^2} \frac{d^n}{dx^n} e^{-x^2}\\
&=& (-1)^n\left[\frac{d}{dx}-2x\right]^n  \nonumber \\
&=&2^n e^{-A} e^{A}
\left[x- \frac{1}{2}\frac{d}{dx}\right]^n
e^{-A}e^{A}\,1\quad.
\eea
The above expression, by suitable manipulations, leads to
\bea \nonumber
H_n(x)= 2^n \exp(-\frac{1}{4} \frac{d^2}{dx^2})\,x^n \quad.
\eea

This shows the procedure to obtain the exponential form of the
solutions, starting from the known RF for a given solution.

\subsection{Generating Functions}

Below, we outline the method of getting the generating functions
(GF) taking Laguerre and Chebyshev as examples. The advantage of
having an exponential form for the solution comes out naturally
through these examples.

\noindent \underline{{\bf GF for the Laguerre Polynomial.}}

Defining the generating function as \be g(x,t) =
\sum^{\infty}_{n=0} L^0_n(x)t^n  \ee and substituting the
expression for $ L^0_n(x)$ in the above equation, we get, \bea
\nonumber
g(x,t) &=& \sum^{\infty}_{n=0} \frac{(-1)^n}{n!} e^{-B}x^n t^n  \\
\nonumber
        &=& e^{-B} e^{-xt} \\
        &=& \left[ 1 - B + \frac{B^2  }{2!} - \frac{B^3}{3!} + \cdots
\right] e^{-xt} \quad.
\eea Action $B$ on $e^{-xt}$ is easy to compute: \be g(x,t) =
e^{-xt} \left[ 1 - (xt^2-t) + \frac{1}{2!} (x^2t^4 - 4xt^3 + 2t^2)
- \cdots \right] \ee This series can be summed and written in the
compact form \bea \nonumber g(x,t) = \frac{\exp[-xt/(1-t)]}{(1-t)}
\quad, \eea $g(x,t)$ is the well-known GF for $L^0_n(x)$. The
procedure outlined above can be straightforwardly extended to
include, $\alpha \ne 0$ cases also.

\noindent \underline{{\bf Chebyshev type II}}

We consider type II Chebyshev polynomial for illustration. The
solution of Chebyshev type II DE can be written as\be U_n(x)= 2^n
\exp{\left[-\frac{1}{2(D+n+2)}\frac{d^2}{dx^2} \right]}x^n \quad,
\ee and its GF is defined as, \be g(x,t)=
\sum^{\infty}_{n=0}U_n(x)t^n \quad. \ee Substituting the
expression for $U_n(x)$ in the above equation, we obtain, \bea
\nonumber g(x,t)&=& \sum^{\infty}_{n=0}2^n
\exp{\left[-\frac{1}{2(D+n+2)}\frac{d^2}{dx^2} \right]}x^n t^n
\quad,\\ \nonumber\\ &=& 1+2xt+(4x^2-1)t^2+ \cdots\quad,\eea which
can be summed to yield the GF as \bea \nonumber g(x,t)=
\frac{1}{1-2xt+t^2}\quad. \eea This procedure for obtaining GF can
be generalized to other orthogonal polynomials. It is clear that
the exponential form of the solutions enables one to find the GFs
straightforwardly.

Generalized GFs find application in the construction of coherent
and squeezed states \cite{fer1,nieto1}. It is worth mentioning
that the exponential form of the Hermite polynomials has been
connected with the Gauss transform \cite{nieto2}. Our results
provide analytic expressions for the Gauss transform, with
appropriate parameter value, of various polynomials. Construction
of generalized coherent states, associated with these polynomials
is currently under progress and will be reported elsewhere.

\section{Conclusions.}

In conclusion, the solution of a wide class of liner differential
equations, which can be cast into a form, where a function of the
Euler operator and constants separates from the rest, can be
written in a closed form, which makes the algebraic properties of
the solution space transparent. Explicit examples dealing with
confluent hypergeometric, hypergeometric and generalized
hypergeometric equations were analyzed, where, not only novel
expressions connecting the solution space, with the space of the
monomials were written down, but also, utilized for unravelling
the dynamical symmetryies underlying the solution spaces.

The fact that, {\em a priori} no assumption was made about the
symmetry of the equation understudy, makes this approach
attractive. Although, we have analyzed here well-known examples,
an exhaustive study reveals that the present approach extends to a
host of other functions and polynomials. Some of these are,
Meijer's G-Function, Struve, Lomel, Anger, Weber, Bessel
functions, Gegenbauer, Neumann's, Jacobi, Schl\"afli, Whittaker,
Chebyshev polynomials. These functions and polynomials manifest in
diverse branches of physics and mathematics. The novel expressions
for the solutions presented here will help in unravelling various
properties of these functions and polynomials.

Here, we have only briefly mentioned about the multivariate cases
dealing with correlated systems.  The solution space of these
equations have rich symmetry and they are connected with random
matrices which find application in diverse areas \cite{random}.
Hence, a deeper analysis of the the problem is also warranted.
Furthermore, the connection of this approach  with the Gauss
transform can be used for constructing generalized coherent
states. It is worth mentioning that, the exponential form of the
Hermite polynomial has already found application in the
construction of coherent and squeezed states \cite{fer1,nieto1}.

Apart from finding exact solutions, the present approach can also
be utilized for finding approximate solutions to differential
equations \cite{atre}. Some of these works are currently under
progress and we hope to report the findings in the near future.

\noindent {\bf Acknowledgements}

\noindent We acknowledge useful discussions with Profs. N.
Mukunda, V. Srinivasan, R. Jagannathan, S. Chaturvedi and R.
Sridhar. T.S. thanks UGC (India) for providing financial support
through the JRF scheme.

\end{document}